# Properties of wideband resonant reflectors under fully conical light incidence

Yeong Hwan Ko, Manoj Niraula, Kyu Jin Lee, and Robert Magnusson[*]

Department of Electrical Engineering, University of Texas at Arlington, Box 19016, Arlington, Texas 76019, USA
[*]*magnusson@uta.edu*

**Abstract:** Applying numerical modeling coupled with experiments, we investigate the properties of wideband resonant reflectors under fully conical light incidence. We show that the wave vectors pertinent to resonant first-order diffraction under fully conical mounting vary less with incident angle than those associated with reflectors in classical mounting. Therefore, as the evanescent diffracted waves drive the leaky modes responsible for the resonance effects, fully-conical mounting imbues reflectors with larger angular tolerance than their classical counterparts. We quantify the angular-spectral performance of representative resonant wideband reflectors in conic and classic mounts by numerical calculations with improved spectra found for fully conic incidence. Moreover, these predictions are verified experimentally for wideband reflectors fashioned in crystalline and amorphous silicon in distinct spectral regions spanning the 1200-1600-nm and 1600-2400-nm spectral bands. These results will be useful in various applications demanding wideband reflectors that are efficient and materially sparse.

## 1. Introduction

It is well known that the optical response of periodic diffractive lattices depends strongly on the angle of incidence of the illuminating wave. For a simple one-dimensional (1D) grating, for example, if the grating vector lies in the plane of incidence (POI), the two orthogonal polarization states decouple and all diffraction orders generated lie in the same plane. This common geometry allows straightforward derivation of diffraction efficiencies for all propagating waves for s polarization (electric field vector normal to the POI) and independently for the orthogonal p polarization. This arrangement is sometimes called the classic mount. The general case, however, is conic mount, or conic diffraction, in which the wave vectors of the propagating diffraction orders reside on the surface of a cone [1]. The general diffraction problem for a 1D grating under conic incidence was treated in detail by Moharam et al. applying the rigorous coupled-wave analysis method [1].

Numerous papers have been written on the specific properties of resonant gratings under conic incidence. For example, Peng and Morris formulated 2D resonant gratings quantifying, in particular, a polarization-independent narrow-band response at normal incidence [2]. Mizutani et al. considered 2D structures engaging the flexibility in angular incidence to resonate simultaneous TE and TM leaky modes to attain polarization independence; this idea succeeded across small bandwidths [3]. Fehrembach and Sentenac analyzed resonant-grating eigenmodes in 2D devices emphasizing conditions for polarization-independent filtering [4]. In contrast to these works, Lacour et al. treated 1D gratings and showed that polarization independent filtering can be attained with such elements with correctly chosen parameters [5,6]. In particular, results for unpolarized filtering under full-conical incidence were presented showing s/p polarization reflectance overlapping partially while occupying the same central resonance wavelength [5]. By optimizing the filter parameters numerically, much improved results in full-conic incidence were later realized exhibiting near-complete overlap of computed s/p reflection spectra [6]. Niederer et al. presented experimental measurements showing polarization-independent filtering in 1D gratings in full-conic mount; qualitative agreement with theory was found [7]. Finally, Grinvald et al. discussed and compared conic and classic polarization independent filters also providing experimental measurements with reasonable spectral agreement for the orthogonal polarization states [8]. In all these works, a main goal is polarization-independent filtering; these are achieved in theory and experiment to various degrees across small resonance bandwidths [2-8]. Taking a different view, Peters et al. investigated theoretically the angular sensitivity of narrow-band resonant filters under full-conical incidence [9].

In contrast, in this contribution, we consider the properties of wideband resonant reflectors with subwavelength periodicity under full-conic incidence. Comparing classical incidence (POI parallel to the grating vector) with fully conical incidence (POI perpendicular to the grating vector), we quantify the angular variability of the reflectance for selected example devices. It is shown that a given reflector exhibits improved stability against angle under full-conic incidence. This is mainly attributed to the fact that the first-order diffracted wave vectors vary slowly with angle of incidence in the subwavelength regime. To characterize theoretically the angular reflectance spectra of general 1D guided-mode resonance (GMR) wideband reflectors, we apply rigorous coupled wave analysis (RCWA) to zero-contrast grating (ZCG) and high-contrast grating (HCG) structures that were previously designed by particle swarm optimization (PSO) [10]. This is followed by experimental demonstrations in two spectral bands where good agreement with theoretical calculations is found.

**2. Wave vector analysis**

For a simple comparison of the incident-angle dependent diffraction properties in classical and fully conical mounting, we study the wave vectors associated with the first-order diffracted waves in the 1D grating as these drive the resonant leaky modes. Figure 1 displays a schematic illustration of subwavelength gratings in classical and fully conical mounting. Here, the $m^{\text{th}}$–order diffractive wave vectors in the $xy$ plane ($k_{x,m}$ and $k_{y,m}$) are determined by the grating equations [1, 11]

$$\textit{Classical mounting: } k_{x,m} = k_{inc} n_0 \sin\theta_{inc} - m\mathbf{K}, k_{y,m} = 0, \qquad (1)$$

$$\textit{Fully conical mounting: } k_{x,m} = k_{inc} n_0 \sin\theta_{inc}, k_{y,m} = -m\mathbf{K}, \qquad (2)$$

where $k_{\text{inc}}$, $n_0$, $\mathbf{K}$, and $\theta_{inc}$ denote the incident wave vector, refractive index of the cover medium (air in this paper), grating vector with magnitude $K=2\pi/\Lambda$, and incident angle, respectively. As depicted, the projection of $k_{inc}$ in the $xy$ plane is parallel (in classical mounting) or perpendicular (in full conical mounting) to the direction of the grating vector. The resultant diffracted wave vectors follow Eqs. (1) and (2). As the incident wavelength ($\lambda$) is larger than grating period, all non-zero diffraction orders are evanescent. Coupling of evanescent diffraction orders to leaky waveguide modes contributes the resonance effects foundational to these reflectors. For classical incidence in Fig. 1, the first evanescent diffracted waves ($k_{diff,1st}$) propagate forward ($m = -1$) and backward ($m = +1$) along the $x$–axis with different magnitudes. In contrast, full conic incidence generates two first-order diffracted waves symmetric around the $x$–axis [12]. As the incident angle ($\theta_{inc}$) changes, the $k_{diff,1st}$ vectors vary differently in these two mounting arrangements

Figures 2(a) and (b) display the angular variation of the normalized first-order diffracted wave vector ($|\Delta k_{diff,1st}|/K$) as function of the normalized incident wave vector ($k_{inc}/K$) in classical and fully conical incidence, respectively. Here, we define the variation of the first-order diffracted wave ($|\Delta k_{diff,1st}|$) as

$$\left|\Delta k_{diff,1st}\right|/K = \left\|k_{diff,1st}\right| - \left|k_{diff,1st}(\theta_{inc}=0)\right\|/K \qquad (3)$$

$$\textit{Classical mounting: } \left|\Delta k_{diff,1st}\right| = k_{inc} n_0 \sin\theta_{inc}, \qquad (4)$$

$$\textit{Fully conical mounting: } \left|\Delta k_{diff,1st}\right| = \sqrt{\left(k_{inc} n_0 \sin\theta_{inc}\right)^2 + K^2} - K, \qquad (5)$$

Comparing these contour plots, it is clearly seen that fully conical mounting provides a more stable angular response. At $k_{inc}/K = 0$, the $|\Delta k_{diff,1st}|/K$ increases slightly from 0 to 0.06 as $\theta_{inc}$ increases from 0 to 45 °in fully conical mounting. In contrast, under classical incidence, it increases considerably more to 0.36.

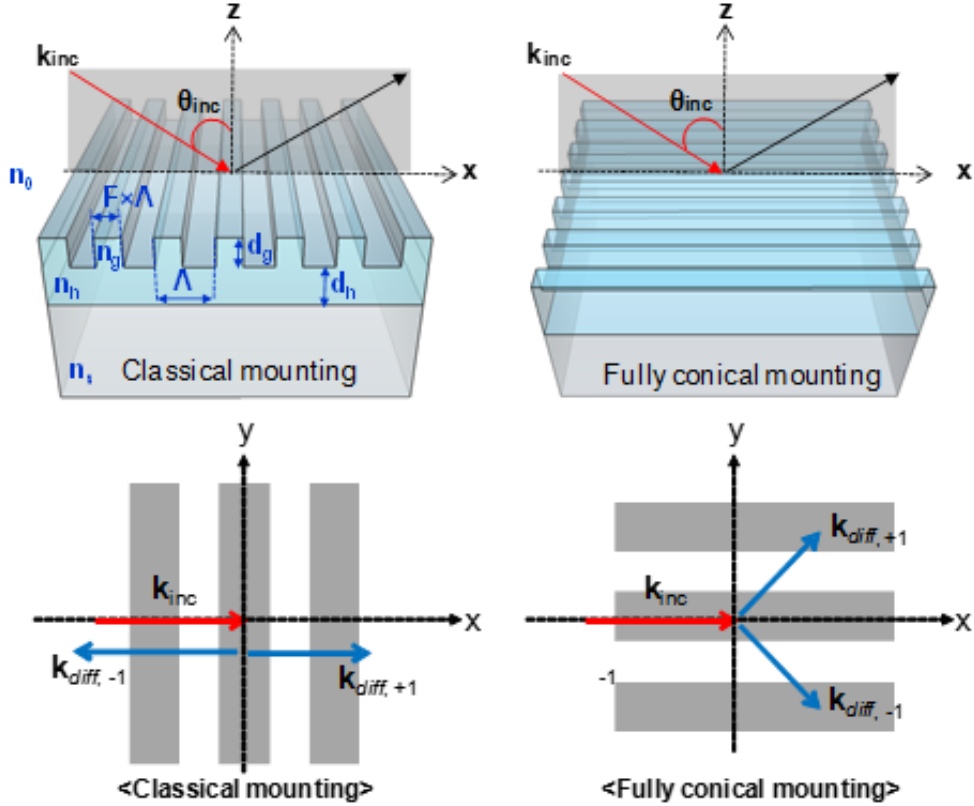

Fig. 1. Schematic illustration of 1D subwavelength gratings in classic and fully conic mounts. The grating parameters are grating period ($\Lambda$), fill factor (F), grating depth ($d_g$), and thickness of the homogeneous sublayer ($d_h$). Here, $n_g$, $n_h$, $n_s$, and $n_0$ denote the refractive indices of the grating, sublayer, substrate and air.

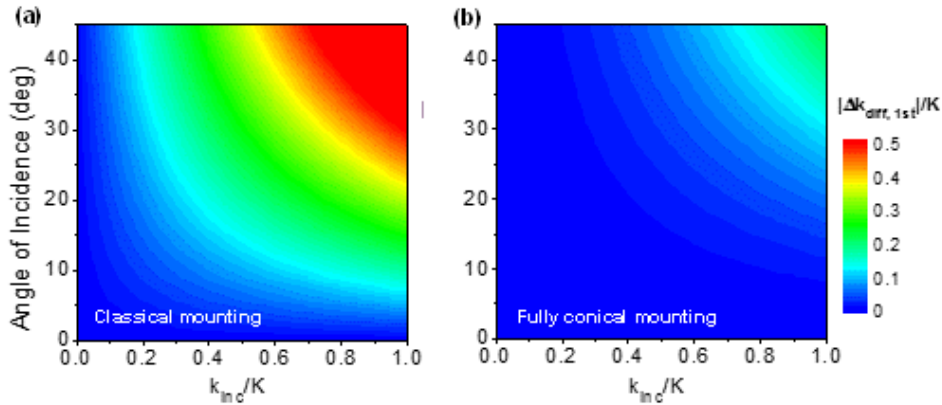

Fig. 2. Variation of the normalized first-order diffracted wave vector ($|\Delta k_{diff,1st}|/K$) as function of angle of incidence ($\theta_{inc}$) and normalized incident wave vector ($k_{inc}/K$). These are calculated in cases of (a) classical and (b) fully conical mounting respectively where $K$ is the grating vector $2\pi/\Lambda$.

### 3. Theoretical high-contrast and zero-contrast grating results

We begin by investigating theoretically the angular tolerance of wideband resonant reflectors under fully conical incidence using previously PSO-optimized 1D grating structures fashioned as high-contrast gratings (HCG) and zero-contrast gratings (ZCG) [13]. Figure 3 shows computed results for the zeroth-order reflectance ($R_0$) as a function of incident angle under TM polarization. The HCG is at (a) classical and (b) fully conical incidence and the ZCG is at (c) classical and (d) fully conical incidence. The HCG and ZCG parameters are {$d_g$ = 493 nm, $\Lambda$ = 786 nm, and F = 0.707} and {$d_g$ = 470 nm, $d_h$ = 255 nm, $\Lambda$ = 827 nm, and F = 0.643}. The refractive indices of the grating ($n_g$), sublayer ($n_h$) and substrate ($n_s$) are 3.48, 3.48, and 1.48, respectively.

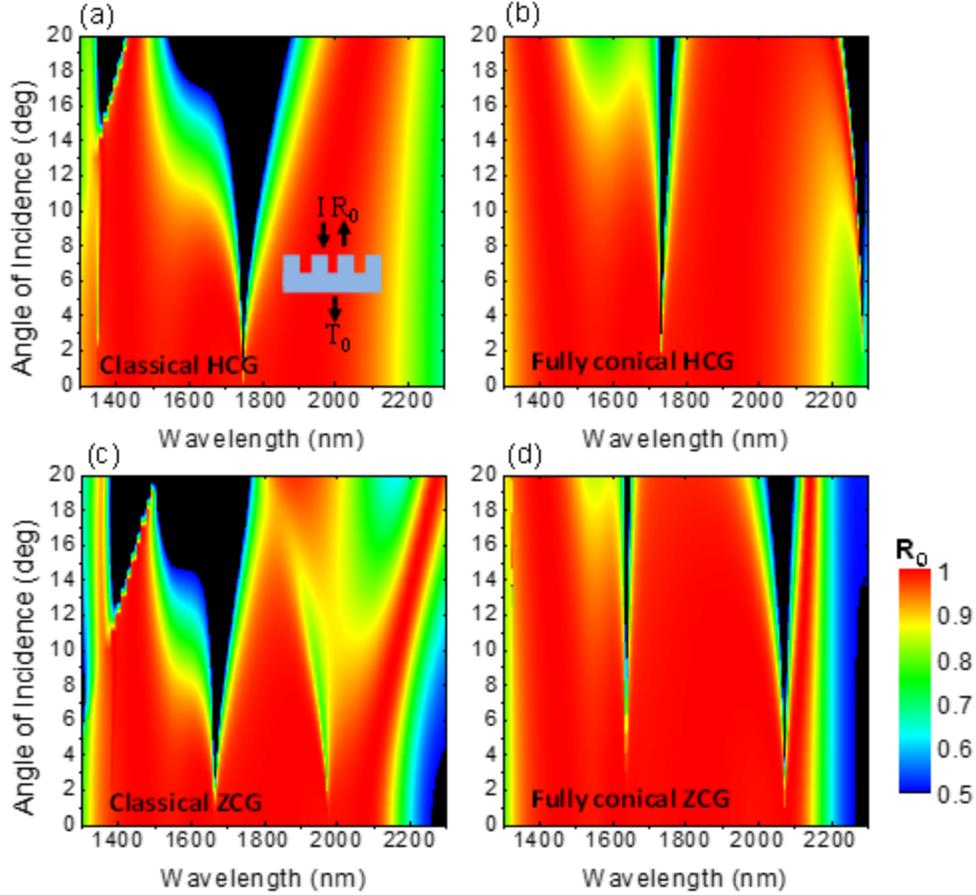

Fig. 3. Calculated zeroth-order reflectance ($R_0$) maps of a HCG GMR reflector at (a) classical and (b) fully conical mounting as a function of angle of incidence. The grating parameters are $d_g$ = 493 nm, $\Lambda$ = 786 nm, and F = 0.707. For the ZCG GMR reflector, the $R_0$ maps are also calculated in cases of (c) classical and (d) fully conical mounting. The grating parameters are $d_g$ = 470 nm, $d_h$ = 255 nm, $\Lambda$ = 827 nm, and F = 0.643. These HCG and ZCG GMR reflector examples were first published in [13] but with $\theta_{inc}$ = 0.

In case of the HCG reflector, in Figs. 3(a) and (b), the reflection band is split at ~1730-1740 nm with increasing incidence angle. However, at fully conical incidence, the two split bands maintain a larger high-reflection area for large incident angles. At $\theta_{inc}$ = 20°, $R_0$ > 95% still covers wide bands ~1823–2216 nm (bandwidth ~393 nm) for fully conic incidence while it covers ~1252–1408 nm (width ~166 nm) for classical incidence. It is notable that the angular tolerance can be improved simply by full conical mounting of the same device. Similarly, in Figs. 3(c) and (d), the ZCG reflector also exhibits more stable reflection bands against incident angle in fully conical mounting. When the angle increases, the reflection band splits into three divided bands that are more robust at fully conical incidence. Particularly, the middle reflection band in Fig 3(d) sustains high reflection exceeding 95% in the range of 1690-1913 nm, which means the acceptance angle for 223 nm bandwidth of $R_0$>95% can be significantly increased from 7° to 20° by mounting the device orthogonally.

For the specific cases in Figs. 3(c) and (d), we characterize the angular variation of first-order diffraction. Figure 4 shows the $k_{diff,1st}$ as a function of incident angle for the ZCG in classical and fully conical mounting. Here, we set the wavelength of the incident light to $\lambda = 1800$ nm for the middle part of the three split bands. When the incident angle increases from 0 to 45° in classical mounting, the first-order diffractive waves propagate forward such that $k_{+1}$ largely increases its magnitude from 7.597 μm$^{-1}$ to 5.129 μm$^{-1}$. Also, the magnitude of the backward $k_{-1}$ diffractive vector varies from 7.597 μm$^{-1}$ to 10.066 μm$^{-1}$. On the other hand, in fully conical mounting, these components vary from 7.597 μm$^{-1}$ to 7.988 μm$^{-1}$. This stable magnitude of first-order diffraction enables retention of guided-mode resonance under high angular variation.

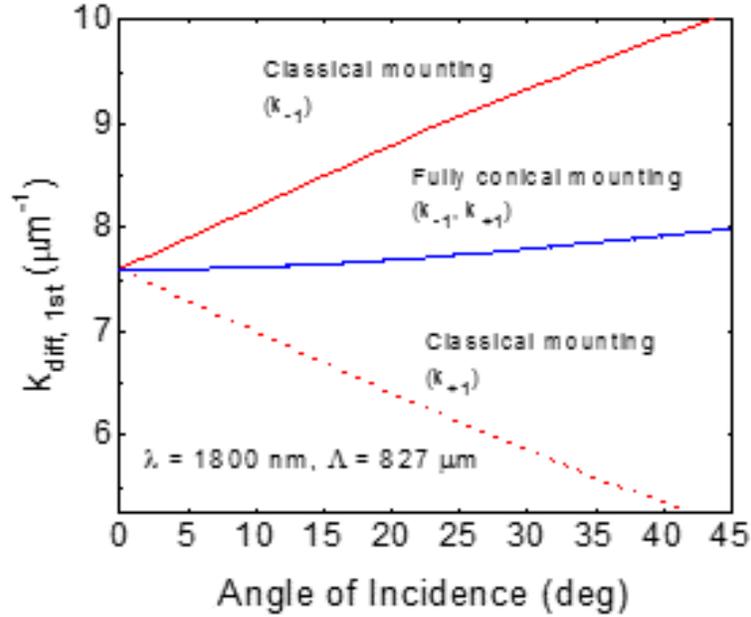

Fig. 4. Plots of the first-order diffracted wave vector ($k_{diff,1st}$) as a function of incident angle for a ZCG with $\Lambda = 827$ nm and the incident light wavelength $\lambda=1800$ nm under classical (red lines) and fully conical (blue lines) mounting.

**4. Experimental results and discussion**

We fabricate a ZCG reflector to operate in the near-IR band based on a silicon-on-quartz (SOQ) platform. The ZCG geometry is defined as period $\Lambda = 560$ nm, Si grating depth $d_g = 330$ nm, homogeneous Si-layer thickness $d_h = 190$ nm, and grating fill factor F = 0.63. These parameters are obtained through the PSO method aiming for a wideband reflection response under TM polarization in the 1200-1500 nm wavelength range, as shown in Fig 5(a). There, calculated $R_0 > 99\%$ covers ~180 nm wide band from 1259 to 1439 nm. In the simulation, we account for material dispersion and extinction in crystalline-Si (c-Si) using values reported in [14]. For dispersion in quartz, we use the material library built into the RSoft code. To experimentally quantify the tolerances of this high-efficiency reflection band against classical and fully conical angles of incidence, we fabricate the ZCG mirror using a commercially available SOQ wafer (Shin-Etsu Chemical Co, Ltd) with 520-nm-thick c-Si film on quartz substrate. Implementing holographic lithography methodology [15] we expose and develop UVN-30, a negative photoresist, to create 1D resist mask on the c-Si film. We reactive-ion etch the c-Si film in SF$_6$ + CF$_4$ ambient and remove the residual resist by ashing in O$_2$. A scanning electron micrograph (SEM) of a representative ZCG is shown in Fig. 5(b); it is seen that the fabricated device exhibits a uniform grating profile with minimal surface blemishes. Estimated fabricated device parameters from the SEM are the same as the design parameters.

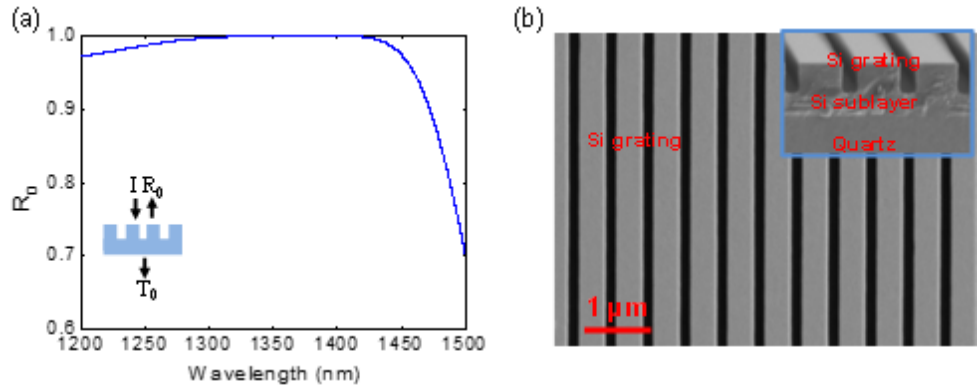

Fig. 5. (a) RCWA simulated TM spectral response of a ZCG reflector with PSO optimized parameters $\Lambda$ = 560 nm, $d_g$ = 330 nm, $d_h$ = 190 nm, and F = 0.63 at normal angle of incidence. (b) SEM showing top-view image of a fabricated ZCG reflector. Inset figure shows cross-sectional SEM profile

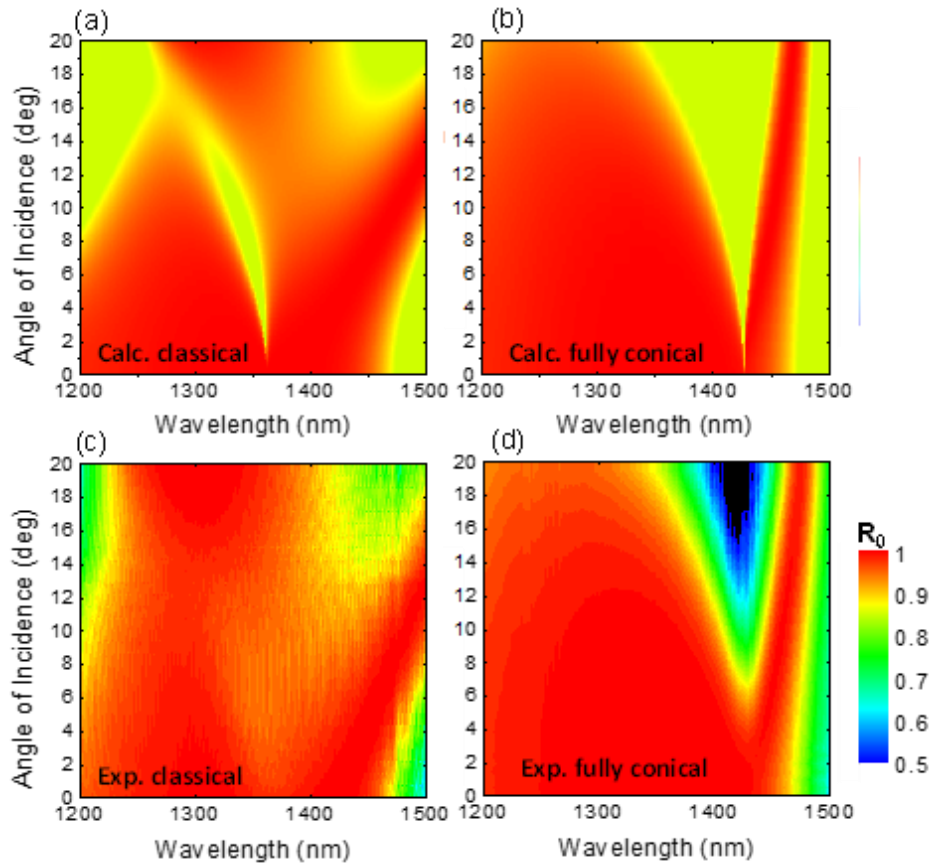

Fig. 6. Calculated $R_0$ maps of a ZCG resonant reflector in (a) classical and (b) fully conical mounting as a function of the angle of incidence. The grating parameters are $d_g$ = 330 nm, $d_h$ = 190 nm, $\Lambda$ = 560 nm, and F = 0.63. Experimentally, the measured $R_0$ maps are presented in cases of (c) classical and (d) fully conical mounting.

We measure the zero-order transmittance ($T_0$) through the reflector by simply normalizing the transmitted signal with the input-beam intensity and proceed to approximate $R_0$ as $1-T_0$. We operate exclusively in the subwavelength regime, where only zero-order waves can propagate. Moreover, at $\lambda > 1200$ nm, the extinction coefficient of c-Si is $< 10^{-6}$ [14], and it becomes virtually lossless at that point. For this case of low-loss, scatter-free media and subwavelength regime operation, we conclude that $R_0 = 1-T_0$ is a reasonable approximation. For optical measurements, we use a super-continuum light source and a near-IR optical spectrum analyzer. We measure $R_0$, through the $1-T_0$ approximation, for various angles of classical and fully conical incidence. RCWA simulated $R_0$ maps for classical and fully conical incidence under TM polarization are shown in Figs. 6(a) and 6(b), respectively. Corresponding measurement results are shown in Figs. 6(c) and 6(d). As for the examples in Fig 3, we see that fully conical incidence provides a wider high-reflectance band than classical incidence. For instance, at $\theta_{inc} = 5°$, $R_0 > 99\%$ covers two separates bands from ~1266–1346 nm and ~1391-1438 nm for classical incidence. The total coverage band extends 127 nm. For fully conical incidence, $R_0 > 99\%$ covers ~1252–1408 nm (for a total of 156 nm) band at the same angle. The RCWA simulated $R_0$ maps in Figs. 6(a)-(b) are in reasonable agreement with the experimental results presented in Figs. 6(c)-(d).

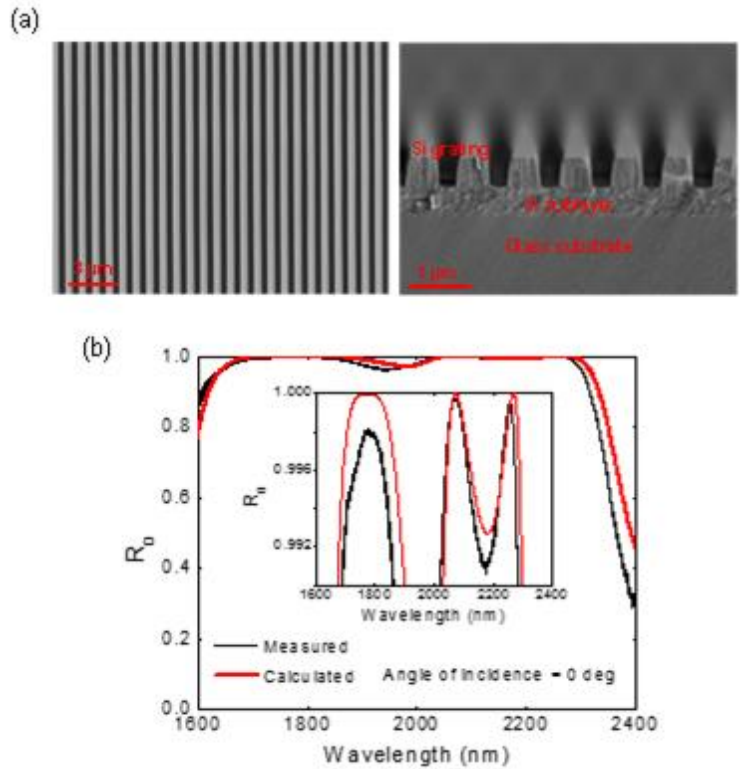

Fig. 7. A ZCG wideband reflector for the longer-wave near-IR region (1680–2300 nm). (a) SEM top and cross-sectional views of the a-Si grating on a glass substrate and (b) measured (black line) and calculated (red line) $R_0$ spectra of the corresponding device. The grating parameters are $d_g = 565$ nm, $d_h = 410$ nm, $\Lambda = 858$ nm, and F = 0.55.

We design and fabricate another ZCG reflector operating at longer wavelengths in the near-IR region (1600–2400 nm). The device is now made with amorphous Si (a-Si) sputtered onto a glass substrate subjected to similar fabrication steps as the c-Si device just described. The experimental grating parameters of $d_g = 565$ nm, $d_h = 410$ nm, $\Lambda = 858$ nm, and F = 0.55 are estimated from the SEM in Fig. 7(a). As shown in the perspective view, the 1D ZCG forms on the insulating layer without any major defects. In Fig. 7(b), the measured and calculated $R_0$ spectra under TM polarization exhibit excellent quantitative agreement. In the measurements, the $R_0 = 1-T_0$ approximation is applied. The high-reflectance band exceeding 95% in $R_0$ covers a spectral range of 1637–2318 nm. As can be seen in the inset plot, this 683-nm-wide reflection band is formed by a combination of three resonance peaks.

Figure 8 shows the calculated and measured angle-dependent $R_0$ maps of the corresponding device under TM polarization; calculated maps are shown in (a) and (b) and measured maps are shown in (c) and (d) for classical and fully conical incidence, respectively. The experimental results show quantitative agreement with the simulation results. As the incident angle is increased, the wide reflection band splits to form three resonance peaks, as noted in the inset of Fig. 7(b). For classical mounting, in Fig. 8(a), only one reflection band (~1700–1850 nm) survives above 13°. However, under fully conical mounting in Fig. 8(b), the three reflection bands still maintain high reflection up to 19°. This trend is also observed in the experimental results shown in Figs. 8(c) and 8(d). At $\theta_{inc} = 16°$, $R_0 > 90\%$ covers a single band across ~1718–1925 nm, thus extending across 207 nm, for classical incidence. In contrast, $R_0 > 90\%$ distributes over three divided bands for fully conical incidence, namely 1762–1636 nm (width ~126 nm), 1799–2073 nm (width ~274 nm), and 2131–2325 nm (width ~94 nm). In addition, the fully conical mounting provides a large acceptance angle of 18° for $R_0 > 90\%$ at longer wavelengths of 2154–2325 nm (width ~171 nm).

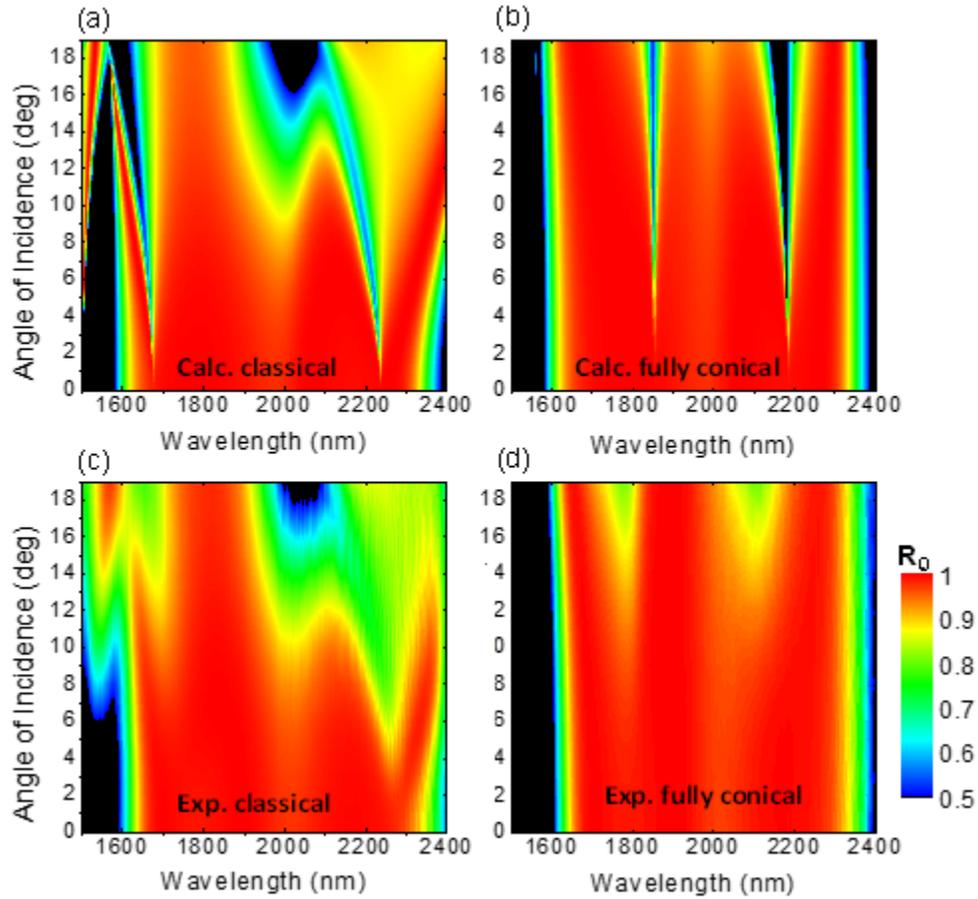

Fig. 8. Calculated $R_0$ maps for the ZCG reflector in Fig. 6 under (a) classical and (b) full conical mounting as a function of angle of incidence. Measured $R_0$ maps for a ZCG reflector with the same parameters for (c) classical and (d) fully conical mounting. The angular region is limited to the range of 0°–19° as exclusively zero-order waves propagate in this spectral/angular domain.

## 5. Conclusions

In conclusion, we show that wideband resonant reflectors under full conical incidence provide improved stability relative to variation in angle of incidence. Hence, the angular tolerance of these reflectors can be improved by simply mounting the same device orthogonally. This angular stability can be explained by slow variation of the wave vectors associated with first-order diffraction against angle of incidence in the subwavelength regime. This impacts reflector response as the resonant leaky modes sustaining the high reflectance are driven by these wave vectors via phase matching. Simulation results show that both HCG and ZCG structures exhibit the improved angular tolerance in full conic mount. We confirm the theoretically predicted angular tolerance of the ZCG-reflector variety by fabricating two devices that operate in distinct spectral regions. In both cases, full conical mounting offers improved angular tolerance with large acceptance angles for various reflection bands. The theoretical and experimental results presented here will be useful for various optical-engineering applications in which wideband, efficient, and materially-sparse reflectors are needed.

**Acknowledgements**

This research was supported, in part, by the National Science Foundation under award no. IIP-1444922. The authors thank Shin-Etsu Chemical Co, Ltd., Japan, for providing the SOQ wafers.

**References**

1. M. G. Moharam, E. B. Grann, D. A. Pommet, and T. K. Gaylord, "Formulation for stable and efficient implementation of the rigorous coupled-wave analysis of binary gratings," J. Opt. Soc. Am. A **12**(5), 1068–1076 (1995).
2. S. Peng and G. M. Morris, "Resonant scattering from two-dimensional gratings," J. Opt. Soc. Am. A **13**(5), 993–1005 (1996).
3. A. Mizutani, H. Kikuta, K. Nakajima, and K. Iwata, "Nonpolarizing guided-mode resonant grating filter for oblique incidence," J. Opt. Soc. Am. A **18**(6), 1261–1266 (2001).
4. A. L. Fehrembach and A. Sentenac, "Study of waveguide grating eigenmodes for unpolarized filtering applications," J. Opt. Soc. Am. A **20**(3), 481–488 (2003).
5. D. Lacour, J-P. Plumey, G. Granet, and A. Mure-Ravaud, "Resonant waveguide grating: analysis of polarization independent filter," Opt. Quantum Electron. **33**(4), 451–470 (2001).
6. D. Lacour, G. Granet, J.-P. Plumey, and A. Mure-Ravaud, "Polarization independence of a one-dimensional grating in conical mounting," J. Opt. Soc. Am. A **20**(8), 1546–1552 (2003).
7. G. Niederer, W. Nakagawa, H. P. Herzig, and H. Thiele, "Design and characterization of a tunable polarization-independent resonant grating filter," Opt. Express **13**(6), 2196–2200 (2005).
8. E. Grinvald, T. Katchalski, S. Soria, S. Levit, and A. A. Friesem, "Role of photonic bandgaps in polarization-independent grating waveguide structures," J. Opt. Soc. Am. A **25**(6), 1435–1443 (2008).
9. D. W. Peters, R. R. Boye, and S. A. Kemme, "Angular sensitivity of guided mode resonant filters in classical and conical mounts," Proc. SPIE **8633**, 86330W (2013).
10. R. Eberhart and J. Kennedy, "Particle swarm optimization," Proc. IEEE Conference on Neural Networks **4**, 1942–1948 (1995).
11. Sakoolkan Boonruang, *Two-dimensional Guided Mode Resonant Structures for Spectral Filtering Applications* (Ph.D. dissertation, University of Central Florida, 2007).
12. A.-L. Fehrembach, D. Maystre, and A. Sentenac, "Phenomenological theory of filtering by resonant dielectric gratings," J. Opt. Soc. Am. A **19**(6), 1136–1144 (2002).
13. R. Magnusson, "Wideband reflectors with zero-contrast gratings," Opt. Lett. **39**(15), 4337–4340 (2014).

14. M. A. Green, "Self-consistent optical parameters of intrinsic silicon at 300 K including temperature coefficients," Sol. Energy Mater. Sol. Cells **92**(11), 1305–1310 (2008).
15. W. W. Ng, C.-S. Hong, and A. Yariv, "Holographic Interference Lithography for Integrated Optics," IEEE Trans. Elect. Dev. **25**(10), 1193–1200 (1978).